\documentclass[reprint,amssymb,amsmath,superscriptaddress,aps]{revtex4-2}
\pdfoutput=1

\usepackage{graphicx,soul}
\usepackage{bm}
\usepackage{bbm}
\usepackage{amsmath}
\usepackage{amsthm}
\usepackage{amssymb}
\usepackage{tabularx}
\usepackage{gensymb}
\usepackage{amssymb}
\usepackage{amsmath}
\usepackage{graphicx}
\usepackage{dcolumn}
\usepackage[usenames, dvipsnames]{color}
\usepackage[caption=false]{subfig}
\usepackage{bm}
\usepackage[colorlinks=true,linkcolor=blue,citecolor=blue,urlcolor=blue,breaklinks=true]{hyperref}
\newcommand{\blu}{\color{blue}}
\newcommand{\blk}{\color{black}}

\definecolor{ngreen}{rgb}{0.2,0.6,0.2}
\definecolor{amethyst}{rgb}{0.6, 0.4, 0.8}

\definecolor{golden}{rgb}{0.75,0.6,0.15}

\DeclareMathOperator*{\argmin}{arg\,min}

\DeclareMathAlphabet\mathbfcal{OMS}{cmsy}{b}{n}
\newcommand{\past}[1]{\overleftarrow{#1}}

\newcommand{\bra}[1]{\langle #1|}
\newcommand{\ket}[1]{|#1\rangle}

\newcommand{\ro}[1]{\left( #1 \right)}
\begin{document}



\title{Error-tolerant witnessing of divergences in classical and quantum statistical complexity}

\author{Farzad \surname{Ghafari}}
\email{f.ghafari@griffith.edu.au}
\affiliation{Centre for Quantum Dynamics, Griffith University, Yuggera Country, Brisbane, Queensland 4111, Australia}
\author{Mile Gu}
\email{mgu@quantumcomplexity.org}
\affiliation{Nanyang Quantum Hub, School of Physical and Mathematical Sciences, Nanyang Technological University, Singapore 639673, Republic of Singapore}
\affiliation{Centre for Quantum Technologies, National University of Singapore, 3 Science Drive 2, Singapore, Republic of Singapore}
\author{Joseph Ho}%
\affiliation{Centre for Quantum Dynamics, Griffith University, Yuggera Country, Brisbane, Queensland 4111, Australia}
\author{Jayne~Thompson}%
\affiliation{Horizon Quantum Computing, 29 Media Cir., Singapore 138565}
\affiliation{Centre for Quantum Technologies, National University of Singapore, 3 Science Drive 2, Singapore, Republic of Singapore}
\author{Whei Yeap Suen}%
\affiliation{Centre for Quantum Technologies, National University of Singapore, 3 Science Drive 2, Singapore, Republic of Singapore}
\author{Howard M. Wiseman}%
\email{h.wiseman@griffith.edu.au}
\affiliation{Centre for Quantum Dynamics, Griffith University, Yuggera Country, Brisbane, Queensland 4111, Australia}
\affiliation{Centre for Quantum Computation and Communication Technology (Australian Research Council), Griffith University, Brisbane, 4111, Australia}
\author{Geoff J. Pryde}%
\email{g.pryde@griffith.edu.au}
\affiliation{Centre for Quantum Dynamics, Griffith University, Yuggera Country, Brisbane, Queensland 4111, Australia}

\date{\today}

\begin{abstract}
How much information do we need about a process' past to faithfully simulate its future? The statistical complexity is a prominent quantifier of structure for stochastic processes. Quantum machines, however, can simulate classical stochastic processes while storing significantly less information than their optimal classical counterparts. This implies qualitative divergences between classical and quantum statistical complexity. Here, we develop error-tolerant techniques to witness such divergences, enabling us to account for the inevitable imperfections in realising quantum stochastic simulators with present-day quantum technology. We apply these tools to experimentally verify the quantum memory advantage in simulating an Ising spin chain, even when accounting for experimental distortion. This then leads us to observe a recently conjectured effect, the \textit{ambiguity of simplicity}---the notion that the relative complexity of two different processes can depend on whether we model the process using classical or quantum means of information processing. 

\end{abstract}

\maketitle




\section{Introduction} 

A key measure of complexity for stochastic processes in time is the minimal amount of past information a model needs to generate a statistically faithful simulation of its conditional future~\cite{Crutchfield1989,Grassberger1986}. Thus a process that changes completely random from one moment to the next, or one that remains unchanged over all times, are both as simple as possible in this sense. 
No information about their past is needed to reproduce, statistically, their future, since the future does not depend on the past. By contrast, a simple alternating sequence, $\ldots 101010 \ldots $, is more complex, since one needs to remember the value of the last bit to know if the next bit will be $0$ or $1$. This measure, formalised as the \textit{statistical complexity}, has seen use in diverse settings ranging from financial analytics to many-body physics~\cite{Crutchfield1989,Shalizi2001,Crutchfield2009}. 

The concept of statistical complexity can be generalised by allow quantum machines to store the information about the past process, and to generate the future process~\cite{Gu2012}. The so-called quantum statistical complexity can exhibit drastic qualitatives from its classical counterpart
~\cite{Gu2012,Aghamohammadi2016,Binder2018}. Specifically, the memory requirements of quantum models can exhibit more favourable scaling~\cite{garner2017provably,Aghamohammadi2017}, such that processes that require ever growing classical memory to simulate may only require a quantum memory of bounded dimension~\cite{elliott2020extreme}. In addition, the relative order of what is complex can differ --- a parametrised family of processes can become more complex to model classically, but less complex to model quantum mechanically ---  a phenomena termed the  \textit{ambiguity of simplicity}~\cite{Aghamohammadi2016}. The reduced complexity of quantum machines can have significant operational benefit, enabling quantum-enhanced rare-event sampling~\cite{aghamohammadi2018extreme}, time-series analysis~\cite{Elliott2018} and more thermally efficient means of stochastic simulation~\cite{Cabello2016}. More fundamentally, they illustrate the striking feature of quantum information's capacity to reshape our perceptions of what is complex.


However, experimental verification of these quantum--classical divergences remains challenging. One of the issues is to ensure that the demonstration is robust to experimental imperfections. Notably, the execution of quantum models assumes the availability of perfect unitary operations. Any environmental noise would result in a distorted model --- one whose memory costs and predictive statistics would differ from the ideal. In previous proof-of-principle realizations of quantum models, these effects were neglected when comparing the resulting quantum memory requirements of such models with optimal classical limits~\cite{ghafari2019dimensional}. Thus, two issues remained: (1) the quantum models may have simulated a slightly different process that might has significantly reduced classical statistical complexity; and (2) the noise accumulation in the quantum model may induce growing memory requirements if the simulation continued to operate {\em ad infinitum}.

Our goal here is to address these issues by developing a method to explicitly characterise and account for such imperfections. We demonstrate these techniques by building an optical quantum simulator for an iconic system in statistical physics --- the one-dimensional (1D) Ising spin chain~\cite{Book-Pathria1972}. We conclusively demonstrate its reduced quantum statistical complexity by simulating the Ising system for a range of temperatures. In addition, we identify regions where there are pairs of temperatures, $T_A$ and $T_B$, for which the classical \blk and quantum notions of relative complexity are reversed -- providing the first experimental demonstration of the \textit{ambiguity of simplicity}.

\section{Materials and methods}
\subsection{Framework} Statistical complexity quantifies the structure of a one-dimensional classical stochastic process. For ease of discussion, we will take this one dimension as a time dimension. Then statistical complexity measures the minimum amount of information a model must carry about the process's past in order to or generate, its possible futures faithfully (i.e. with the correct statistics). 

To be more precise, we imagine a dynamical system that emits discrete-valued outputs $x_t$---instances of random variables $X_t$---at discrete times $t \in \mathbb{Z}$. The output string, $\cdots, X_{-1}, X_0, X_1, X_2,  \cdots$, is a stochastic process, described by a joint probability distribution, $\mathcal{P} = P(\overleftarrow{X},\overrightarrow{X})$. Here $\overleftarrow{X}=  \cdots,  X_{-1}, X_0 $ and $ \overrightarrow{X}= X_1, X_2,\cdots$  respectively represent the past and future strings at time $t=0$. Any faithful model of the process must replicate this behaviour. That is, for each observed past $\overleftarrow{x}$, the model must provide a systematic means of initializing a suitable machine $\Xi$ in some state $\epsilon(\overleftarrow{x})$, such that repeated application of a systematic action 
$M$ on $\Xi$ sequentially generates $x_1$, $x_2 \ldots$ governed by the conditional future $P(\overrightarrow{X}|\overleftarrow{X} = \overleftarrow{x})$. Here $\epsilon(\cdot)$ is referred to as the \emph{encoding function}, capturing precisely how the model encodes the past within its memory. 

Given a faithful model of a stochastic process $\mathcal{P}$, the amount of past information the model stores is given by $S(\Xi)$, the Shannon entropy of $\Xi$. The \emph{statistical complexity} of this process then corresponds to the value of $S(\Xi)$ when minimized over all faithful models~\cite{Crutchfield1989,Shalizi2001}. Traditionally, this minimization assumed that $\Xi$ is classical, whereby the minimum is achieved by $\epsilon$-machines~\cite{Crutchfield1989,Crutchfield1994}. This involves assigning each possible past $\overleftarrow{x}$ to an appropriate memory state $S_i = \epsilon(\overleftarrow{x})$ for some $i\in \{1,\dots,N\}$, known as a causal state, such that two pasts are assigned to the same state if and only if their conditional futures statistics coincide. The $\epsilon$-machine then uses a classical memory $\Xi$ that associates each memory state $S_i$ with a different physical configuration. The statistical complexity is thus defined by ${C_\mu} =  - \sum_{i = 1}^N{{p_i}}\,\mathrm{log}_2\,{p_i}$, where 
\begin{equation}
    p_i = \sum_{\overleftarrow{x}} P(\overleftarrow{X}=\overleftarrow{x}) 
\delta_{S_i,\epsilon(\overleftarrow{x})}
\end{equation}
is the steady-state probability that the model is in causal state $S_i$. The identification of such causal states and the associated memory cost to store them have been used for stochastic analysis in diverse contexts, including many-body physics, neural spike trains, financial analytics and chaotic dynamics~\cite{Crutchfield2012,Crutchfield2009,Crutchfield2003,Crutchfield1989}.

Despite the fact that the stochastic process here is completely classical, there are benefits to be gained by considering quantum modelling, in which the machine $\Xi$ is a quantum physical system. In particular, this has the potential of further reducing memory cost~\cite{Gu2012,Mahoney2016,Binder2018}. Such $q$-machines can encode relevant past information \textit{coherently}---replacing each $S_i$ with a quantum state $\left| {{S_i}} \right\rangle$. The resulting memory then has entropy
\begin{equation}
C_q =  - \mathrm{Tr}(\rho \log_2 (\rho)),
\label{eq:Cq}
\end{equation}
where $\rho  = \sum_i^N {{p_i}} \left| {{S_i}} \right\rangle \left\langle {{S_i}} \right|$. In this case, repeated application of a suitable quantum instrument  
on a system $\Xi$ in state $\ket{S_i}$ generates correct conditional future statistics, i.e. samples from $P(\overrightarrow{X}|\past{x} \in S_i)$, while leaving $\Xi$ in the state $\ket{S_j}$, for some $j\in\{1,\dots,N\}$,  
appropriate to the new past, $\past{x}x_1$. 
In general, $C_q \leq C_\mu$, and these two quantities can exhibit drastically different qualitative and scaling behaviour~\cite{Aghamohammadi2016,garner2017provably,elliott2020extreme}.

\subsection{Simulating the Ising system}  Determination of the true quantum statistical complexity requires minimization of $C_q$ over all valid choices of $\left| {{S_i}} \right\rangle$. This optimization is generally non-trivial, but has been achieved for the 1D Ising system~\cite{Suen2017}. This, coupled with the clear physical relevance of Ising systems, make them an ideal candidate for witnessing the differences between quantum and classical statistical complexity. 

\begin{figure}
	\centering
	\includegraphics[width=1\linewidth]{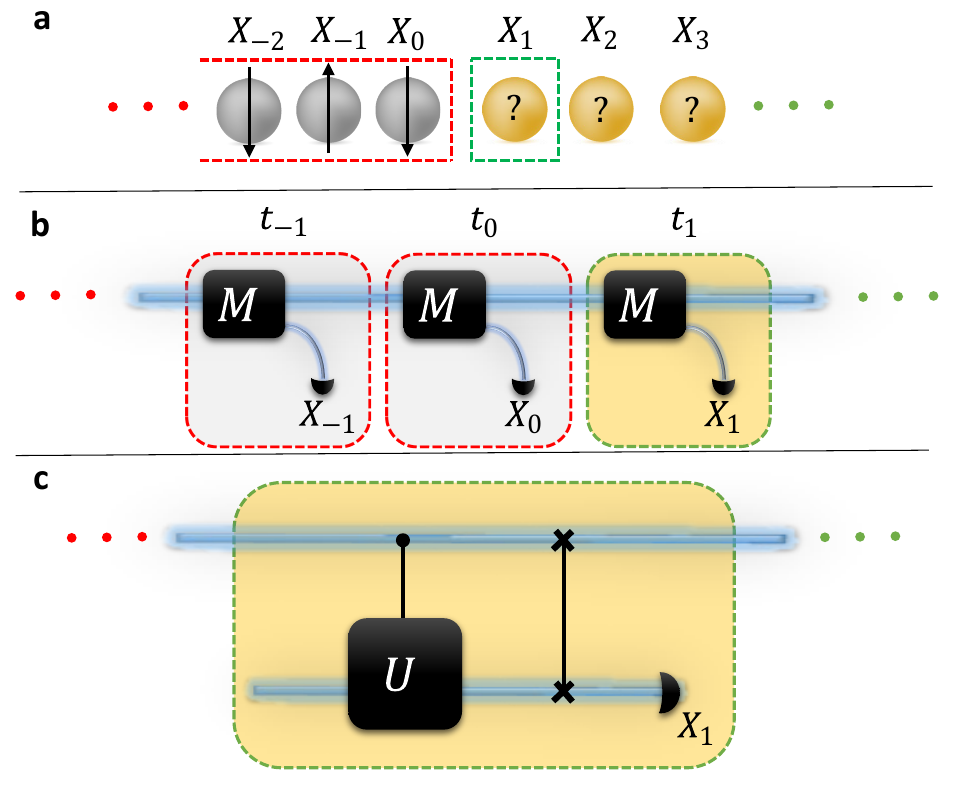}
	\caption{Concept and workings of a quantum machine to simulate Ising statistics. 
	\textbf{(a)} A 1D Ising system is a spatial chain of spins, each represented as stochastic binary variables $X_n$. This system is scanned from left to right and the observed spins (gray with red border), to the left of the current place (yellow with green border) correspond to past. Spins to its right (yellow) correspond to the future. It is their probabilistic behaviour which is to be simulated, given the past observations. 
	\textbf{(b)} Simulation of the spin statistics in discrete time. 
	The $\epsilon$-machine encodes causal states in a physical system, the memory. At each time step $t_n$, the action of $\textit{M}$ on the causal state,  generates $X_n$ with the appropriate statistics, while updating the memory in the relevant causal state for the next step. \textbf{(c)}) For a quantum $\epsilon$-machine, each simulator step is implemented using an entangling gate (a quantum controlled-unitary followed by a swap gate) between the input causal state encoded in the memory and an ancilla. The readout provides the classical outcome and updates the memory for the next step in the relevant quantum causal state.}
	\label{fig:concept}
\end{figure}

Specifically, we consider a 1D Ising system describing an infinite spatial chain of spins with nearest-neighbour interactions. At any non-zero temperature, this string must be described statistically. 
In this case we can replace the method for simulation over a series of discrete times (described above) with a simulation over discrete spatial sites, corresponding to scanning the system spatially (e.g. left to right) through spin locations $n$. In this way, the ``past'' corresponds to all spins to the left of the current position and the ``future'' corresponds to all spin sites to the right~\cite{Crutchfield1997}, and we can use our technique to examine the spatial statistics as desired. This is illustrated in Fig.~\ref{fig:concept}. 

For the Ising system, the energy function is
\begin{equation}
H(\overleftarrow{x}, \overrightarrow{x}) = - \sum\limits_n \ro{  J{x_n}{x_{n + 1}} + B{x_n}},
\end{equation}
where $J$ is the coupling parameter, $B$ the magnetic field, and ${x_n} \in \{  - 1,1\} $ is the spin at site $ n $. For each configuration, at temperature $T $, the joint probability distribution is given by the Boltzmann distribution~\cite{Book-Pathria1972}. We use natural units for temperature ($ k_{\rm B}\blk=1 $) and take the coupling $ |J| $ to be the unit of energy so that $ |J|=1 $ and $ T $ and $ B $ are dimensionless.

This process has two causal states, $\{S_i\}_{i=0:1}$, with encoding function $\epsilon(\overleftarrow{x})$ that identifies any two pasts with coinciding $x_0$~\cite{Crutchfield1997}. The corresponding $\epsilon$-machine then stores these two causal states within its memory. At each time-step, it operates according to the transition probabilities $\Gamma_{ij}(J,B,T)$, which represent the probability a simulator in state $S_i$ will transit to $S_j$ while emitting output $j$ (see Appendix~\ref{App1}).

Meanwhile, the provably optimal quantum model has quantum causal states
\begin{subequations}
	\begin{eqnarray}
	\left| {{S_0}} \right\rangle & =& \sqrt{\Gamma_{00}}  \left| 0 \right\rangle \, + \sqrt{ \Gamma_{01}} \left| 1 \right\rangle \label{eq:casualstatea}\\
	\left| {{S_1}} \right\rangle & =& \sqrt {\Gamma_{10}} \left| 0 \right\rangle \, + \sqrt {\Gamma_{11}} \left| 1 \right\rangle \label{eq:casualstateb},
	\end{eqnarray}
	where
	\label{eq:casualstate}
\end{subequations}
$\left| 0 \right\rangle$ and $\left| 1 \right\rangle$ are orthogonal qubit states. Statistics for the subsequent spin can then be generated by interacting this memory with an ancilla initial set in state $\ket{0}$ via the controlled-unitary $U_C$, satisfying
\begin{subequations}
	\begin{eqnarray}
	U_C |S_0\rangle| 0 \rangle  & =& \sqrt {{\Gamma_{00}}} | 0 \rangle| {S_0}\rangle  + \sqrt {{\Gamma_{01}}} | 1 \rangle| {S_1},\rangle \\
	U_C | S_1 \rangle | 0 \rangle  & = &\sqrt {{\Gamma_{10}} }| 0 \rangle| {S_0}\rangle  + \sqrt {{\Gamma_{11}}} | 1 \rangle| {S_1}\rangle,
	\end{eqnarray}
	\label{eq:gate}
\end{subequations}

\noindent followed by a measurement of the ancilla qubit. In general, $\ket{S_0}$ and $\ket{S_1}$ are mutually non-orthogonal. Thus the quantum statistical complexity $C_q$ is less than $C_\mu$.
\noindent
%


\subsection{Accounting for Imperfection}  
\begin{figure}
	\centering
\includegraphics[width=1\linewidth]{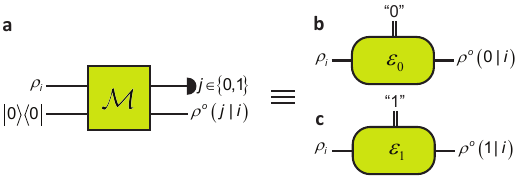}
	\caption{Process map of the physical gate. 
	\textbf{(a)}) Due to experimental imperfections, a slightly non-unitary operation $ \mathcal{M} $ is implemented. The ${\cal M}$ can be decomposed into two single-qubit completely-positive maps \textbf{(b)}) $ \mathcal{E}_0 $ and \textbf{(c)}) $ \mathcal{E}_1 $, conditional on the 0 and 1 ancilla measurement outcomes, respectively.}
	\label{fig:map}
\end{figure}

To witness the reduced quantum statistical complexity $C_q$ experimentally, the direct method would be to synthesize the quantum circuit representing the quantum model. However, small experimental imperfections~\cite{Rohde2005} mean that instead of an ideal controlled-unitary gate, a more general transformation $\mathcal{E}$ is implemented. The situation is depicted in Fig.~\ref{fig:map}. For the ideal process, if causal state $\rho_i =|S_i\rangle \langle S_i| $ is the input, then the output of the circuit is $\rho_j = |S_j\rangle \langle S_j|$ when outcome  $j \in \{0,1\}$  is obtained at the measurement. To allow for non-idealities we describe this by a map $\mathcal{M}$, that does not necessarily preserve purity. 

Because one of the input qubits (the ancilla) is always prepared in a fixed state (ideally $\ket{0}\bra{0}$), in place of the two-qubit completely positive (CP) trace-preserving map ${\cal M}$ 
it is more convenient to use a one-qubit {\em instrument}. This comprises two one-qubit CP trace-decreasing maps $\mathcal{E}_0$ and $\mathcal{E}_1$, acting on the memory qubit state $\rho_i$, depending on whether the measurement outcome was 0 or 1. The result is the output state $\rho^o(j|i)$ multiplied by the probability $\wp^o(j|i)$ with which it occurs: 
\begin{equation}
   \wp^o(j|i) \rho^o(j|i)=\mathcal{E}_j(\rho_i) 
\end{equation}
For an ideal experiment, $\wp^o(j|i)=\Gamma_{ij}$ and $\rho^o(j|i)=\rho_j$. But experimental imperfections can lead to deviations in both of those equations. This \blk creates two complications: (1) the quantum device will simulate a process different from the one it is aimed to simulate (the Ising chain, in our case) and (2) the output states from one step of the simulator, $\rho^o(j|i)$, is now \textit{not}, in general, equal to either of the possible input causal states. Thus, its memory cost will generally not be $S(\rho_i)$.

Previous experiments were proof-of-principle quantum simulators, and thus mostly ignored these imperfections for purposes of characterizing $C_q$~\cite{Palsson2017,ghafari2019interfering,ghafari2019interfering}. The entropy was taken to be the entropy of the output states after one step of the process, and there was no consideration of how small errors in this might cause the simulator to ultimately diverge in future predictions. 
Here, we aim to build quantum simulators that illustrate that $C_q$ and $C_\mu$ behave very differently even when we account for such imperfections.



We begin by developing a method of doing the quantum--classical comparison fairly in the presence of imperfections. 
We first identify \blk target parameters $B$, $J$ and $T$ and tune our experiment to realize a quantum $\epsilon$-machine that approaches these parameters. Next, we perform quantum process tomography~\cite{Book-Nielsen2010, White2007,Bongioanni2010} of the circuit to obtain the $ \mathcal{E}_0 $ and $ \mathcal{E}_1 $ maps. From these maps, we find the states and transition probabilities $ \{\rho^{m}_i,\Gamma^{m}_{ij} \} $ ($m$ is for ``maps''), which describe the two-state quantum machine most closely corresponding to these maps. That is, the approximate equality $ \mathcal{E}_j( {\rho _i^m} )\approx \Gamma_{ij}^m{\rho _j^m}$ is as close as possible to an equality. Here, closeness is defined in terms of trace distance (see Appendix~\ref{App2}) 
We call the $ \rho^{m}_i $  the \textit{fixed-point states}. The stationary state of our machine is then $ \rho^{m}={p^{m}_0}{\mkern 1mu} {\rho_0^{m}}{\mkern 1mu}  + {\mkern 1mu} {p^{m}_1}{\mkern 1mu} {\rho^{m}_1} $, where $ p^{m}_0 = \Gamma^{m}_{10}/({\Gamma^{m}_{10}+\Gamma^{m}_{01}})$ and $ p^{m}_1=1-p^{m}_0 $. Since the $\Gamma_{ij}$ are  theoretical functions of $T$ and $B$ (equation A2 of Appendix~\ref{App1}), we can numerically invert the equation $ \Gamma_{ij}(J=1,B^m,T^m) =  \Gamma^{m}_{ij}$, to find the $T^m$ and $B^m$ that our real map $\mathcal{E}$ actually implement (see Appendices~\ref{App2} and \ref{App3} for details).

Having determined the fixed-point states of the maps, we then go a further step by using them as input causal states. We collect statistics from our experimental simulator at various nominal values of $T$, for fixed $ B $ and $ J $ (The nominal values are listed in Appendix~\ref{App1}).
The binary outputs of the simulation are found by measuring the ancilla output in the logical basis, and these statistics are used to determine $\Gamma^{s}_{ij}$ (where $s$ denotes a value derived from the statistical output). This yields the corresponding stationary state probabilities $p^s_0=\Gamma^s_{10}/({\Gamma^s_{10}+\Gamma^s_{01}})=1-p^s_1$, and inferred values of $T^s$ and $B^s$. We also define a corresponding stationary state, 
$ \rho^{s}={{p^{s}_0}{\mkern 1mu} {\rho^{m}_0}{\mkern 1mu}  + {\mkern 1mu} {p^{s}_1}{\mkern 1mu} {\rho^{m}_1}}$, that would result from repeated applications of the map with these transition probabilities. The entropy of $\rho^{s}$ thus corresponds to the quantum statistical complexity of the Ising process at inferred parameter values of $T^s$ and $B^s$. A comparison of this to the classical statistical complexity for the same parameter values accounts for all the effects of experimental imperfection.

\section{Results}

\subsection{Experimental Setup} 
\begin{figure}
	\centering
	\includegraphics[width=1\linewidth]{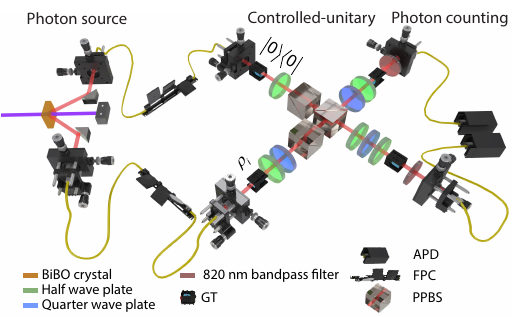}
	\caption{Experimental set-up. Memory and ancilla photonic polarization qubits are generated by spontaneous parametric down conversion. The controlled-unitary operation is based on a controlled-$ Z $ gate~\cite{Langford2005} implemented by partially-polarizing beam splitters (PPBSs) and single-qubit operations implemented by wave plates. GT stands for Glan Taylor prism, and APD for avalanche photo diode. FPC are fiber polarization controllers.}
	\label{fig:setup}
\end{figure}

Here, we implement one complete cycle of the $\epsilon$-machine, comprising memory state preparation, a controlled-unitary operation, and read-out (see Fig.~\ref{fig:setup}\textbf). Unentangled single-photon polarization qubits are produced by degenerate spontaneous parametric down conversion (SPDC). The source was realized using a 410 nm cw pump laser and a BiB${}_3$O${}_6$ (BiBO) crystal cut for type-I phase matching. We use polarization to encode logical states $ \left| 0 \right\rangle=\left| H \right\rangle  $ and $\left| 1 \right\rangle=\left| V \right\rangle $, where $ H $ and $ V $ are horizontal and vertical polarization, respectively. 

The optical simulation circuit is based around single-qubit unitary rotations implemented with wave plates and a nondeterministic linear optics controlled-Z gate using three partially-polarizing beam splitters (PPBSs)~\cite{Langford2005}.  Our desired $U_C$ gate can then be decomposed to these components, operations by noting
$$ U_C  = (I \otimes {V_0})(I \otimes {V_1})(I \otimes H)Z_C{(I \otimes H)^{ - 1}}{(I \otimes {V_1})^{ - 1}}$$, where ${V_0}\left| 0 \right\rangle {\rm{  =  }}\left| {{S_0}} \right\rangle$, $ V_1 $ is the rotation in the $ X-Z $ plane that rotates $ |0\rangle $ to the bisector of $ |S_0\rangle $ and $ |S_1\rangle $, and $ H$ is a Hadamard operation.

  \begin{figure*} [tb]
	\centering
	\includegraphics[width=1\linewidth]{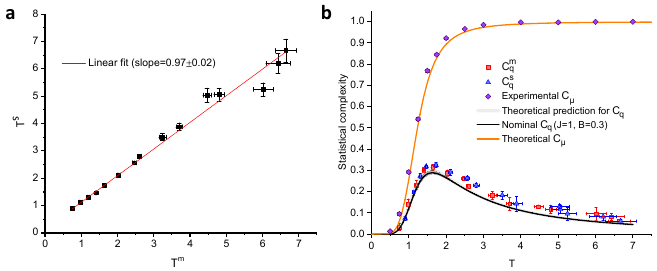}
	\caption{Experimental results, showing the quantum and classical statistical complexity for the ferromagnetic Ising chain with nominal values of $ J=1,\ B=0.3 $ and a range of different temperatures $T$. \textbf{(a)} Implemented temperatures; $ T^s $ versus $ T^m $. \textbf{(b)} Statistical complexity of the $ \epsilon $-machine simulating a 1D Ising spin chain. Error bars are derived from Poissonian photon statistics. The experimentally-determined quantum statistical complexity ${C^{m}_{q}}$ and ${C^{s}_{q}}$ (see text) are plotted against the relevant temperature parameters, $ T^{m} $ and $ T^{s} $ respectively, for the ferromagnetic case with $ J=1, B=0.3 $.
		 The black curve shows the theoretical estimation of $C_{q} $ for the nominal values of $J $, $B$ and $T$. The grey shaded region (close to the black curve) shows the estimated theoretical bound, using the Ising model parameter distribution corresponding to the processes implemented by our  reconstructed  experimental map $\mathcal{ M}$ (see Appendix~\ref{App4}). The orange curve and purple data points are the theoretical and experimental classical statistical complexity $ C_{\mu} $, respectively, where the error bars are from Poissonian photon statistics. Note that the classical curve is monotonic while the quantum curve is not, giving rise to the phenomenon of ambiguity of simplicity~~\cite{Aghamohammadi2016}.}
	\label{fig:result}
\end{figure*}

The polarization qubits are measured using wave plates and avalanche photodiodes. Quantum state and process tomography are implemented using the methods in Ref.~\cite{White2007}. 
The quantum process tomography 
enables us to obtain the specific maps $\mathcal{E}_0$ and $\mathcal{E}_0$ and thus obtain a full description of our experimental quantum model. Following the methodology outlined above to account for these imperfections, we find the best-fit `model'  parameter values $T^m$ and $B^m$, and fixed-point states $\rho^m_i$,  and then the `statistical' values $T^s$,  $B^s$, and $\rho^s$ from the outcomes of applying the experimental map on the $\rho^m_i$.

\subsection{Results}  

In our experiment we consider an ensemble of target Ising systems that we want to simulate, represented by various nominal values of $T$, with fixed $ B=0.30 $ and $ J=1 $. To clarify our procedure, we first walk through how we process the data for an example member of that ensemble. The example is the Ising systems with the nominal temperature value $T=1.50$. 
We prepared the apparatus, as close as we can achieve, to simulating the Ising model with $\{J=1, B=0.30, T=1.50\}$.  
 From  quantum process tomography, we found the states and transition probabilities $  \rho^{m}_i,\Gamma^{m}_{ij} $. By numerically inverting the equation $ \Gamma_{ij}(J=1,B^{m},T^{m}) =  \Gamma^{m}_{ij}$, the values of $T^m =1.40\pm 0.02$ and $B^m=0.27\pm 0.01$  were found. Here, the errors are obtained from Poissonian photon
statistics~\cite{fox2006quantum}.  Then, we prepared $\rho^m_i$ as the input states and collected the statistics which gave  $\Gamma^{s}_{ij}$. The revised values of $T^s =1.46\pm 0.03$ and $B^s=0.28\pm 0.01$ were obtained \blk by inverting equations $ \Gamma_{ij}(J=1,B^{s},T^{s}) =  \Gamma^{s}_{ij}$. We then repeated the same process for the other nominal $T$ values in our ensemble (see Appendix~\ref{App1} for a list of nominal values). 

As expected, there is not perfect agreement between the target parameters and the experimentally obtained parameters $T^{m(s)}$ and $B^{m(s)}$. However, for the ensemble, the mean values of $B^{m}$ and $B^{s}$ were $ 0.29 \pm 0.02 $ and $ 0.28\pm 0.05 $, respectively. The errors here are standard deviation of the mean of the ensemble. The closeness to the constant nominal value of $B=0.3$ shows that the processes implemented in the lab are not far from the target processes. Thus the variation of $B$ from its  nominal value of $0.3$ is negligible, and we consider the behaviour of the system as the temperature is varied (see Appendices~\ref{App3} and \ref{App4}). We plot $ T^{m} $ versus $ T^{s} $ in Fig.~\ref{fig:result}\text{a}, demonstrating that the two methods of reconstructing the process agree well. This agreement is evidence that the simulated machines are close to the nominal ones, and have fixed-point states close to satisfying $\mathcal{E}_j(\rho^{m}_i) = \Gamma^{m}_{ij} \rho^{m}_j$.


Using equation~(\ref{eq:Cq}), we can calculate the quantum statistical complexity both for the theoretical prediction of the steady state based on the tomographically determined model of the simulator, $C_q^m = C_q (\rho^m)$, and for the statistical results  from  implementing the experiment, $C_q^s = C_q (\rho^s)$. We plot these values of the quantum statistical complexity at each temperature ($T^m$ and $T^s$ respectively) in Fig.~\ref{fig:result}b for the case where we are simulating a ferromagnetic chain with nominal values of $B=0.3$ and $J=1$. We observe that $ C^{m}_{q} $ and $ C^{s}_{q} $ lie close to the estimated theoretical range of statistical complexity values (see Appendix~\ref{App4}). The slight discrepancy between the $m$ and $s$ values primarily arises from small repeatability errors in the experimental simulator settings, and from the fact that the calculated fixed point states do not exactly satisfy $\mathcal{E}_j(\rho^{m}_i) = \Gamma^{m}_{ij} \rho^{m}_j$ for $j={0,1}$ (see Appendix~\ref{App3}).

For comparison, we also implement the classical $ \epsilon $-machine using the same experimental set-up. In this case, $ \left| {{S_0}} \right\rangle  = \left| 0 \right\rangle  $ and $\left| {{S_1}} \right\rangle  = \left| 1 \right\rangle $, and future statistics are generated based on introducing classical randomness~\cite{Palsson2017}. Since the states are orthogonal, they do not inherently contain transition probabilities---we implement these probabilities by preparing orthogonal states in an ensemble of experiments with occupation fractions equal to the probabilities $ p_0 $ and $ p_1 $, respectively. Results for the classical $\epsilon$-machine are overlaid on Fig.~\ref{fig:result}b, and lie close to the theoretical prediction.

\subsection{Ambiguity of Simplicity}  
\begin{figure}
	\centering
	\includegraphics[width=1\linewidth]{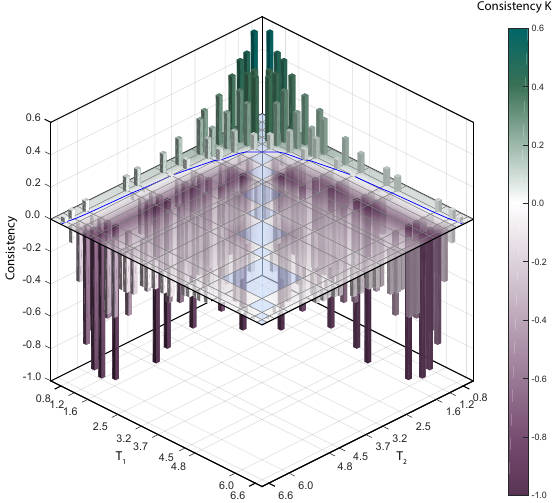}
	\caption{Consistency graph (equation~(\ref{eq:agree}b)), for the experimental data, ${C^{m}_{q}}$ and experimental $ C_\mu $, in part \textbf{a}. For $ -1<K<0 $ the the quantifiers of complexity have ambiguous ordering and for $ 0<K<1 $ they are unambiguous. The pale shading in the plane $ K=0 $ represents a projection of the experimental-result bars onto the plane, and together with the blue curve, demonstrates the boundary between regions of ambiguity and consistency. The pale blue squares in the plane $ K=0 $ indicate the areas where no experimental data exist.}
	\label{fig:consistency}
\end{figure}

An interesting question is whether \textit{relative} simplicity is an intrinsic property of the systems being modelled, not of the models. That is, how does the notion of relative simplicity survive the transition from a classical to a quantum description~\cite{Aghamohammadi2016}? Consider two Ising systems, A and B, with different temperatures $ T_\mathrm{A}$ and $T_\mathrm{B}$. If, in the classical regime, $ C^\mathrm{A}_{\mu} < C^\mathrm{B}_{\mu}$ (which means that A is simpler than B) and, in the quantum regime, $ C^\mathrm{A}_q < C^\mathrm{B}_q$, then there is \textit{consistency} between the two classes for processes A and B. However, if the quantum model reverses its ranking compared with the classical perspective, we have the \textit{ambiguity of simplicity}~\cite{Aghamohammadi2016}. The basic question, ``Which process is simpler?'' no longer has a well-defined answer. To mathematically describe this phenomenon, we define
\begin{subequations}
	\begin{eqnarray}
r(T _1,T _2) &=& \frac{C_q(T _1) - C_q(T _2)}{C_{\mu}(T _1) - C_{\mu}(T _2)}\\\nonumber
K({T _1},{T _2})&=& \text{Sign}(r({T _1},{T _2}))\times\\
&&\min \{ |r(T_1,T_2)|,1/|r(T_1,T_2)|\}.	
	\end{eqnarray}
	\label{eq:agree}
\end{subequations}
\noindent Here $K$ is the degree of consistency. For $-1<K<0$, there is ambiguity according to the definition above, and for $0<K<1$, the models are consistent. The magnitude $|K|\in [0,1]$ gives an indication of the degree of consistency or discrepancy. In Fig.~\ref{fig:consistency}, we construct a diagram that compares all pairs of processes at different temperatures $ T_1 $ and $ T_2 $. As can be seen, the notion of relative physical simplicity, capturing which system needs less memory to simulate, depends on the models used for simulation, i.e. we observe an ambiguity.


\section{Discussion and conclusion}

Here, we  simulated an Ising system  using a quantum  processor that can drastically reduces memory requirements about the system's past, suggesting that complexity exhibits significantly different behaviour in the quantum regime. This involved devising a method to compare quantum and classical notions of statistical measures of complexity that accounts for for experimental imperfections. Moreover, we provide the first experimental witness for ``ambiguity of simplicity''~\cite{Aghamohammadi2016}--- a peculiar feature where the relative order of what is more complex differs depending on whether we allow quantum simulators.

Our work opens a number of interesting future avenues. One observation is that our measure of memory cost is entropic, in line with the most convention in classical literature. Thus, its operational meaning is mostly reflected in the i.i.d regime, corresponding to most costs when simulating a large number of stochastic processes simultaneously. In near-term devices, an alternative single-shot measure using memory dimension may be more relevant. Recent theory and experiments work indicate a quantum advantage in simulation persists and can also scale in this single-shot regime~\cite{elliott2020extreme,ghafari2019dimensional}. Our methods presented here could thus enable a more robust experimental demonstration of such a single-shot simulation advantage. Indeed, the ambiguity of simplicity has not been demonstrated, not within such single-shot regimes, providing further impetus. Meanwhile, there are several other related tasks where quantum devices have the potential at an advantage in memory-limited regimes. Examples include quantum clocks, adaptive agents, and solutions of certain promise problems~\cite{yang2019accuracy,budroni2021ticking,thompson2017using,elliott2021quantum,tian2019experimental}, all of which provide potential candidates for theoretical investigation and experimental realization.

\section{Acknowledgments}

We thank Raj B. Patel for useful discussions and help in performing the experiment. The authors also thank Nora Tischler for discussions and comments on the manuscript. This work was supported by the National Research Foundation of Singapore Fellowship No. NRF-NRFF2016-
02, the FQXi R-710-000-146-720
Grant “Are quantum agents more energetically efficient at
making predictions?” from the Foundational Questions Institute and Fetzer Franklin Fund (a donor-advised fund of Silicon Valley Community Foundation) and the Australian Research Council (project no. DP160101911). H.M.W.'s contribution was funded under the Centre of Excellence Grant CE170100012 (Centre for Quantum Computation and Communication Technology). F.G. and J.H. acknowledge support by the Australian Government Research Training Program (RTP) scholarship.






\begin{appendix}

\section{Ising model}\label{App1}
Different Ising systems may be specified by different $ T, J $ and $ B $. We choose the nominal values of  $\{J=1, B=0.3\}$ as an example of the ferromagnetic regime, and simulate the chain for a range of different nominal temperatures
\begin{equation} \label{Tvalues}
    T\in\{0.75, 1, 1.25, 1.5, 1.75, 2.25, 2.75, 3, 4, 5, 6, 8, 10, 12, 14\}.
\end{equation} 
These are used to calculate nominal values of $\Gamma_{ij}(J,B,T)$ and to realize the causal states defined in equation (3) in the main text. The transition probabilities are given by~\cite{Thesis-Feldman1998}:
	
	\begin{subequations}
		\begin{eqnarray}
	{\Gamma _{00}}& = &{{e^{\frac{{ B + J}}{T}}}}/{D}, \\
	{\Gamma _{01}}& = &1-\Gamma_{00},\\
	{\Gamma _{11}}& = &{{e^{\frac{{ -B + J}}{T}}}}/{D},\\
	{\Gamma _{10}} &=& 1-\Gamma_{11},
		\end{eqnarray}
		\label{eq:TransitionProbs}
	\end{subequations}
\noindent, where $ D = \exp ({J}{T})\cosh({B}{T}) + \sqrt {\exp ( - {2J}{T}) + \exp({2J}{T}){\sinh^2}({B}{T})}$.

\section{Fixed-point states} \label{App2}
In the ideal case defined in equation (\ref{eq:gate}) in the main text, if we get measurement outcome $ j $ with probability $ \Gamma_{ij} $, then $ \mathcal{E}_j(\rho_i) \blu = \blk \Gamma_{ij} \rho_j $.  (Here, the causal state $\rho_i =|S_i\rangle \langle S_i| $ is the input, $\rho_j = |S_j\rangle \langle S_j|$ is the output state of the circuit, and $ \mathcal{E}_0 $ and $ \mathcal{E}_1 $ are the experimentally-implemented maps which are characterized through quantum process tomography performed on the one-qubit process~\cite{White2007,Bongioanni2010}.) However, in practice, a slightly different (but very close) output state $\rho^o(j|i)$ is obtained: it turns out that $\rho^o(j|i)\ne \Gamma_{ij} \rho^o(j|i) $, motivating a theoretical question: ``Given map $ \mathcal{E} $, can we find $ \Gamma_{ij} $ and $ \rho_i $ (for $ i=0,1 $) such that $\mathcal{E}_j(\rho_i) = \Gamma_{ij} \rho_j, $ \textit{exactly}, for $j={0,1}$?''. Experimental tests indicate that the answer to this question is generally ``no'', but it can be close. Instead, we find the best solution for $ \Gamma_{ij} $ and $ \rho_i $ with $ i=0,1 $, as
\begin{equation}\label{eq:7}
\{ {\rho^{m}_i},{\Gamma^{m}_{ij}}\}  = \argmin_{\{ {\rho _i},{\Gamma_{ij}}\}} \left[\sum\limits_{i,j = 0,1} {\left\| {{\mathcal{E} _j}\left( {{\rho _i}} \right) - {\Gamma_{ij}}{\rho _j}} \right\|}  \right],
\end{equation}
where $\argmin_x [f(x)] $ means the value of the arguments $x$ that minimize the function $ f(x) $, and  $\left\| \ldots \right\| $
is the trace distance~\cite{Book-Nielsen2010}.

\begin{figure}
	\centering
	\includegraphics[width=1\linewidth]{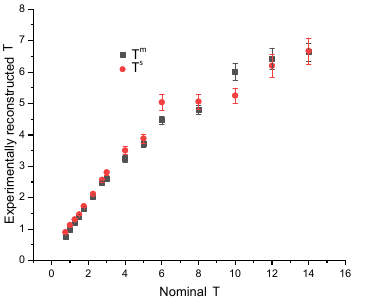}
	\caption{Experimental results for the ferromagnetic case with nominal values of $ J=1,\ B=0.3 $ and a range of different  nominal temperatures $ T $:  $ T^m $ and $ T^s $ versus nominal $ T $.}
	\label{fig:result_app}
\end{figure}

\section{Deviation of implemented Ising parameters from nominal values}\label{App3}
	
The implemented mean value of $ B^m $, $ 0.29 \pm 0.02 $,  is consistent with the nominal value of $ 0.3 $. However, the difference between the nominal value of $T$ and the implemented value of $T$ increases when we aim for high temperatures. As an example, the implemented $ T^m $ is 0.751 when the nominal value is 0.750, while $ T^m $ is 3.237  when the nominal value is 4.000. The implemented values for $ T $ can be seen in Fig.~~\ref{fig:result}{a} of the main text. Also, as it can be seen in Fig.~\ref{fig:result_app}, for values close to $0$, the implemented $T^m$ and $T^s$ versus nominal values almost follow a linear trend, while for higher temperature it starts to saturate.
	
It is possible to obtain an intuitive understanding of why the simulator imperfection increases slightly with increasing $T$ parameter. At low $T$, the quantum causal states are close together (hence the low statistical complexity) but also relatively close to a logical state. At higher $T$, the states are close together but closer to equal superpositions of logical states. Thus gate imperfections that affect coherence (e.g. imperfect or variable phase offsets from the apparatus) are more pronounced on average. Also, the sensitivity of the inversion to find $T^m$ varies with $T$.
 

\section{Theoretical prediction for the statistical complexity of the real simulator}\label{App4}
The simulator maps to an Ising system with temperature $T^{m}$ and magnetic field $B^m$, which vary slightly from the target values in equation~(\ref{Tvalues})
and (constant) $B=0.3$, respectively. While the temperature is the main parameter that varies between different system, the estimated   $B^m$s are not exactly constant. The mean value of $ B^m $ is close to $ 0.30 $, but there is some spread and one can construct a value and an uncertainty band, $B^m \in [0.27 \pm 0.01,0.31 \pm 0.03]$ . For each $B^m$ values, we define an upper bound and a lower bound as ${B^ + } = {B^m} + \sigma $ and ${B^ - } = {B^m} - \sigma $ respectively, where $ \sigma$ is the error bar calculated from Poisson distribution. 
The theoretically predicted quantum statistical complexity for the real simulator is given by $C_q(B \in \{B^+, B^- \},T^{m})$ for $J=1$. The $C_q$ values corresponding to the upper and lower bound resulted in the  grey bounds in Fig.~\ref{fig:result}{b} in the main text.

\end{appendix}


\begin{thebibliography}{30}%
	\makeatletter
	\providecommand \@ifxundefined [1]{%
		\@ifx{#1\undefined}
	}%
	\providecommand \@ifnum [1]{%
		\ifnum #1\expandafter \@firstoftwo
		\else \expandafter \@secondoftwo
		\fi
	}%
	\providecommand \@ifx [1]{%
		\ifx #1\expandafter \@firstoftwo
		\else \expandafter \@secondoftwo
		\fi
	}%
	\providecommand \natexlab [1]{#1}%
	\providecommand \enquote  [1]{``#1''}%
	\providecommand \bibnamefont  [1]{#1}%
	\providecommand \bibfnamefont [1]{#1}%
	\providecommand \citenamefont [1]{#1}%
	\providecommand \href@noop [0]{\@secondoftwo}%
	\providecommand \href [0]{\begingroup \@sanitize@url \@href}%
	\providecommand \@href[1]{\@@startlink{#1}\@@href}%
	\providecommand \@@href[1]{\endgroup#1\@@endlink}%
	\providecommand \@sanitize@url [0]{\catcode `\\12\catcode `\$12\catcode
		`\&12\catcode `\#12\catcode `\^12\catcode `\_12\catcode `\%12\relax}%
	\providecommand \@@startlink[1]{}%
	\providecommand \@@endlink[0]{}%
	\providecommand \url  [0]{\begingroup\@sanitize@url \@url }%
	\providecommand \@url [1]{\endgroup\@href {#1}{\urlprefix }}%
	\providecommand \urlprefix  [0]{URL }%
	\providecommand \Eprint [0]{\href }%
	\providecommand \doibase [0]{http://dx.doi.org/}%
	\providecommand \selectlanguage [0]{\@gobble}%
	\providecommand \bibinfo  [0]{\@secondoftwo}%
	\providecommand \bibfield  [0]{\@secondoftwo}%
	\providecommand \translation [1]{[#1]}%
	\providecommand \BibitemOpen [0]{}%
	\providecommand \bibitemStop [0]{}%
	\providecommand \bibitemNoStop [0]{.\EOS\space}%
	\providecommand \EOS [0]{\spacefactor3000\relax}%
	\providecommand \BibitemShut  [1]{\csname bibitem#1\endcsname}%
	\let\auto@bib@innerbib\@empty
	
	\bibitem{Crutchfield1989}
	\bibinfo{author}{Crutchfield, J.~P.} \& \bibinfo{author}{Young, K.}
	\newblock \bibinfo{title}{Inferring statistical complexity}.
	\newblock \emph{\bibinfo{journal}{Phys. Rev. Lett.}}
	\textbf{\bibinfo{volume}{63}}, \bibinfo{pages}{105} (\bibinfo{year}{1989}).
	
	\bibitem{Grassberger1986}
	\bibinfo{author}{Grassberger, P.}
	\newblock \bibinfo{title}{Toward a quantitative theory of self-generated
		complexity}.
	\newblock \emph{\bibinfo{journal}{Int. J. Theor. Phys.}}
	\textbf{\bibinfo{volume}{25}}, \bibinfo{pages}{907--938}
	(\bibinfo{year}{1986}).
	
	\bibitem{Shalizi2001}
	\bibinfo{author}{Shalizi, C.~R.} \& \bibinfo{author}{Crutchfield, J.~P.}
	\newblock \bibinfo{title}{Computational mechanics: Pattern and prediction,
		structure and simplicity}.
	\newblock \emph{\bibinfo{journal}{J. Stat. Phys.}}
	\textbf{\bibinfo{volume}{104}}, \bibinfo{pages}{817--879}
	(\bibinfo{year}{2001}).
	
	\bibitem{Crutchfield2009}
	\bibinfo{author}{Crutchfield, J.~P.}, \bibinfo{author}{Ellison, C.~J.} \&
	\bibinfo{author}{Mahoney, J.~R.}
	\newblock \bibinfo{title}{Time's barbed arrow: Irreversibility, crypticity, and
		stored information}.
	\newblock \emph{\bibinfo{journal}{Phys. Rev. Lett.}}
	\textbf{\bibinfo{volume}{103}}, \bibinfo{pages}{094101}
	(\bibinfo{year}{2009}).
	
	\bibitem{Gu2012}
	\bibinfo{author}{Gu, M.}, \bibinfo{author}{Wiesner, K.},
	\bibinfo{author}{Rieper, E.} \& \bibinfo{author}{Vedral, V.}
	\newblock \bibinfo{title}{Quantum mechanics can reduce the complexity of
		classical models}.
	\newblock \emph{\bibinfo{journal}{Nat. Commun.}} \textbf{\bibinfo{volume}{3}},
	\bibinfo{pages}{762} (\bibinfo{year}{2012}).
	
	\bibitem{Aghamohammadi2016}
	\bibinfo{author}{Aghamohammadi, C.}, \bibinfo{author}{Mahoney, J.~R.} \&
	\bibinfo{author}{Crutchfield, J.~P.}
	\newblock \bibinfo{title}{The ambiguity of simplicity in quantum and classical
		simulation}.
	\newblock \emph{\bibinfo{journal}{Phys. Lett. A}}
	\textbf{\bibinfo{volume}{381}}, \bibinfo{pages}{1223--1227}
	(\bibinfo{year}{2017}).
	
	\bibitem{Binder2018}
	\bibinfo{author}{Binder, F.~C.}, \bibinfo{author}{Thompson, J.} \&
	\bibinfo{author}{Gu, M.}
	\newblock \bibinfo{title}{Practical unitary simulator for non-markovian complex
		processes}.
	\newblock \emph{\bibinfo{journal}{Phys. Rev. Lett.}}
	\textbf{\bibinfo{volume}{120}}, \bibinfo{pages}{240502}
	(\bibinfo{year}{2018}).
	
	\bibitem{garner2017provably}
	\bibinfo{author}{Garner, A.~J.}, \bibinfo{author}{Liu, Q.},
	\bibinfo{author}{Thompson, J.}, \bibinfo{author}{Vedral, V.} \emph{et~al.}
	\newblock \bibinfo{title}{Provably unbounded memory advantage in stochastic
		simulation using quantum mechanics}.
	\newblock \emph{\bibinfo{journal}{New Journal of Physics}}
	\textbf{\bibinfo{volume}{19}}, \bibinfo{pages}{103009}
	(\bibinfo{year}{2017}).
	
	\bibitem{Aghamohammadi2017}
	\bibinfo{author}{Aghamohammadi, C.}, \bibinfo{author}{Mahoney, J.~R.} \&
	\bibinfo{author}{Crutchfield, J.~P.}
	\newblock \bibinfo{title}{Extreme quantum advantage when simulating classical
		systems with long-range interaction}.
	\newblock \emph{\bibinfo{journal}{Sci. Rep.}} \textbf{\bibinfo{volume}{7}},
	\bibinfo{pages}{6735} (\bibinfo{year}{2017}).
	
	\bibitem{elliott2020extreme}
	\bibinfo{author}{Elliott, T.~J.} \emph{et~al.}
	\newblock \bibinfo{title}{Extreme dimensionality reduction with quantum
		modeling}.
	\newblock \emph{\bibinfo{journal}{Physical Review Letters}}
	\textbf{\bibinfo{volume}{125}}, \bibinfo{pages}{260501}
	(\bibinfo{year}{2020}).
	
	\bibitem{aghamohammadi2018extreme}
	\bibinfo{author}{Aghamohammadi, C.}, \bibinfo{author}{Loomis, S.~P.},
	\bibinfo{author}{Mahoney, J.~R.} \& \bibinfo{author}{Crutchfield, J.~P.}
	\newblock \bibinfo{title}{Extreme quantum memory advantage for rare-event
		sampling}.
	\newblock \emph{\bibinfo{journal}{Physical Review X}}
	\textbf{\bibinfo{volume}{8}}, \bibinfo{pages}{011025} (\bibinfo{year}{2018}).
	
	\bibitem{Elliott2018}
	\bibinfo{author}{Elliott, T.~J.} \& \bibinfo{author}{Gu, M.}
	\newblock \bibinfo{title}{Superior memory efficiency of quantum devices for the
		simulation of continuous-time stochastic processes}.
	\newblock \emph{\bibinfo{journal}{npj Quantum Inf.}}
	\textbf{\bibinfo{volume}{4}}, \bibinfo{pages}{18} (\bibinfo{year}{2018}).
	
	\bibitem{Cabello2016}
	\bibinfo{author}{Cabello, A.}, \bibinfo{author}{Gu, M.},
	\bibinfo{author}{G\"uhne, O.}, \bibinfo{author}{Larsson, J.-r.} \&
	\bibinfo{author}{Wiesner, K.}
	\newblock \bibinfo{title}{Thermodynamical cost of some interpretations of
		quantum theory}.
	\newblock \emph{\bibinfo{journal}{Phys. Rev. A}} \textbf{\bibinfo{volume}{94}},
	\bibinfo{pages}{052127} (\bibinfo{year}{2016}).
	
	\bibitem{ghafari2019dimensional}
	\bibinfo{author}{Ghafari, F.} \emph{et~al.}
	\newblock \bibinfo{title}{Dimensional quantum memory advantage in the
		simulation of stochastic processes}.
	\newblock \emph{\bibinfo{journal}{Physical Review X}}
	\textbf{\bibinfo{volume}{9}}, \bibinfo{pages}{041013} (\bibinfo{year}{2019}).
	
	\bibitem{Book-Pathria1972}
	\bibinfo{author}{Pathria, R.}
	\newblock \emph{\bibinfo{title}{Statistical Mechanics}}
	(\bibinfo{publisher}{Elsevier}, \bibinfo{year}{1972}).
	
	\bibitem{Crutchfield1994}
	\bibinfo{author}{Crutchfield, J.~P.}
	\newblock \bibinfo{title}{The calculi of emergence: computation, dynamics and
		induction}.
	\newblock \emph{\bibinfo{journal}{Physica D}} \textbf{\bibinfo{volume}{75}},
	\bibinfo{pages}{11--54} (\bibinfo{year}{1994}).
	
	\bibitem{Crutchfield2012}
	\bibinfo{author}{Crutchfield, J.~P.}
	\newblock \bibinfo{title}{Between order and chaos}.
	\newblock \emph{\bibinfo{journal}{Nat. Phys.}} \textbf{\bibinfo{volume}{8}},
	\bibinfo{pages}{17--24} (\bibinfo{year}{2012}).
	
	\bibitem{Crutchfield2003}
	\bibinfo{author}{Crutchfield, J.~P.} \& \bibinfo{author}{Feldman, D.~P.}
	\newblock \bibinfo{title}{Regularities unseen, randomness observed: Levels of
		entropy convergence}.
	\newblock \emph{\bibinfo{journal}{Chaos}} \textbf{\bibinfo{volume}{13}},
	\bibinfo{pages}{25--54} (\bibinfo{year}{2003}).
	
	\bibitem{Mahoney2016}
	\bibinfo{author}{Mahoney, J.~R.}, \bibinfo{author}{Aghamohammadi, C.} \&
	\bibinfo{author}{Crutchfield, J.~P.}
	\newblock \bibinfo{title}{Occam’s quantum strop: Synchronizing and
		compressing classical cryptic processes via a quantum channel}.
	\newblock \emph{\bibinfo{journal}{Sci. Rep.}} \textbf{\bibinfo{volume}{6}},
	\bibinfo{pages}{20495} (\bibinfo{year}{2016}).
	
	\bibitem{Suen2017}
	\bibinfo{author}{Suen, W.~Y.}, \bibinfo{author}{Thompson, J.},
	\bibinfo{author}{Garner, A. J.~P.}, \bibinfo{author}{Vedral, V.} \&
	\bibinfo{author}{Gu, M.}
	\newblock \bibinfo{title}{The classical-quantum divergence of complexity in
		modelling spin chains}.
	\newblock \emph{\bibinfo{journal}{Quantum}} \textbf{\bibinfo{volume}{1}},
	\bibinfo{pages}{25} (\bibinfo{year}{2017}).
	
	\bibitem{Crutchfield1997}
	\bibinfo{author}{Crutchfield, J.~P.} \& \bibinfo{author}{Feldman, D.~P.}
	\newblock \bibinfo{title}{Statistical complexity of simple one-dimensional spin
		systems}.
	\newblock \emph{\bibinfo{journal}{Phys. Rev. E}} \textbf{\bibinfo{volume}{55}},
	\bibinfo{pages}{R1239} (\bibinfo{year}{1997}).
	
	\bibitem{Rohde2005}
	\bibinfo{author}{Rohde, P.~P.}, \bibinfo{author}{Pryde, G.~J.},
	\bibinfo{author}{O'Brien, J.~L.} \& \bibinfo{author}{Ralph, T.~C.}
	\newblock \bibinfo{title}{Quantum-gate characterization in an extended hilbert
		space}.
	\newblock \emph{\bibinfo{journal}{Physical Review A}}
	\textbf{\bibinfo{volume}{72}}, \bibinfo{pages}{032306}
	(\bibinfo{year}{2005}).
	
	\bibitem{Palsson2017}
	\bibinfo{author}{Palsson, M.~S.}, \bibinfo{author}{Gu, M.},
	\bibinfo{author}{Ho, J.}, \bibinfo{author}{Wiseman, H.~M.} \&
	\bibinfo{author}{Pryde, G.~J.}
	\newblock \bibinfo{title}{Experimentally modeling stochastic processes with
		less memory by the use of a quantum processor}.
	\newblock \emph{\bibinfo{journal}{Sci. Adv.}} \textbf{\bibinfo{volume}{3}},
	\bibinfo{pages}{1601302} (\bibinfo{year}{2017}).
	
	\bibitem{ghafari2019interfering}
	\bibinfo{author}{Ghafari, F.} \emph{et~al.}
	\newblock \bibinfo{title}{Interfering trajectories in experimental
		quantum-enhanced stochastic simulation}.
	\newblock \emph{\bibinfo{journal}{Nature communications}}
	\textbf{\bibinfo{volume}{10}}, \bibinfo{pages}{1--8} (\bibinfo{year}{2019}).
	
	\bibitem{Book-Nielsen2010}
	\bibinfo{author}{Nielsen, M.~A.} \& \bibinfo{author}{Chuang, I.~L.}
	\newblock \emph{\bibinfo{title}{Quantum Computation And Quantum Information}}
	(\bibinfo{publisher}{Cambridge university press}, \bibinfo{year}{2010}).
	
	\bibitem{White2007}
	\bibinfo{author}{White, A.~G.} \emph{et~al.}
	\newblock \bibinfo{title}{Measuring two-qubit gates}.
	\newblock \emph{\bibinfo{journal}{J. Opt. Soc. Am. B}}
	\textbf{\bibinfo{volume}{24}}, \bibinfo{pages}{172--183}
	(\bibinfo{year}{2007}).
	
	\bibitem{Bongioanni2010}
	\bibinfo{author}{Bongioanni, I.}, \bibinfo{author}{Sansoni, L.},
	\bibinfo{author}{Sciarrino, F.}, \bibinfo{author}{Vallone, G.} \&
	\bibinfo{author}{Mataloni, P.}
	\newblock \bibinfo{title}{Experimental quantum process tomography of
		non-trace-preserving maps}.
	\newblock \emph{\bibinfo{journal}{Phys. Rev. A}} \textbf{\bibinfo{volume}{82}},
	\bibinfo{pages}{042307} (\bibinfo{year}{2010}).
	
	\bibitem{Langford2005}
	\bibinfo{author}{Langford, N.~K.} \emph{et~al.}
	\newblock \bibinfo{title}{Demonstration of a simple entangling optical gate and
		its use in bell-state analysis}.
	\newblock \emph{\bibinfo{journal}{Phys. Rev. Lett.}}
	\textbf{\bibinfo{volume}{95}}, \bibinfo{pages}{210504}
	(\bibinfo{year}{2005}).
	
	\bibitem{fox2006quantum}
	\bibinfo{author}{Fox, A.~M.}, \bibinfo{author}{Fox, M.} \emph{et~al.}
	\newblock \emph{\bibinfo{title}{Quantum optics: an introduction}},
	vol.~\bibinfo{volume}{15} (\bibinfo{publisher}{Oxford university press},
	\bibinfo{year}{2006}).
	
	\bibitem{yang2019accuracy}
	\bibinfo{author}{Yang, Y.}, \bibinfo{author}{Baumg{\"a}rtner, L.},
	\bibinfo{author}{Silva, R.} \& \bibinfo{author}{Renner, R.}
	\newblock \bibinfo{title}{Accuracy enhancing protocols for quantum clocks}.
	\newblock \emph{\bibinfo{journal}{arXiv preprint arXiv:1905.09707}}
	(\bibinfo{year}{2019}).
	
	\bibitem{budroni2021ticking}
	\bibinfo{author}{Budroni, C.}, \bibinfo{author}{Vitagliano, G.} \&
	\bibinfo{author}{Woods, M.~P.}
	\newblock \bibinfo{title}{Ticking-clock performance enhanced by nonclassical
		temporal correlations}.
	\newblock \emph{\bibinfo{journal}{Physical Review Research}}
	\textbf{\bibinfo{volume}{3}}, \bibinfo{pages}{033051} (\bibinfo{year}{2021}).
	
	\bibitem{thompson2017using}
	\bibinfo{author}{Thompson, J.}, \bibinfo{author}{Garner, A.~J.},
	\bibinfo{author}{Vedral, V.} \& \bibinfo{author}{Gu, M.}
	\newblock \bibinfo{title}{Using quantum theory to simplify input--output
		processes}.
	\newblock \emph{\bibinfo{journal}{npj Quantum Information}}
	\textbf{\bibinfo{volume}{3}}, \bibinfo{pages}{1--8} (\bibinfo{year}{2017}).
	
	\bibitem{elliott2021quantum}
	\bibinfo{author}{Elliott, T.~J.}, \bibinfo{author}{Gu, M.},
	\bibinfo{author}{Garner, A.~J.} \& \bibinfo{author}{Thompson, J.}
	\newblock \bibinfo{title}{Quantum adaptive agents with efficient long-term
		memories}.
	\newblock \emph{\bibinfo{journal}{arXiv preprint arXiv:2108.10876}}
	(\bibinfo{year}{2021}).
	
	\bibitem{tian2019experimental}
	\bibinfo{author}{Tian, Y.}, \bibinfo{author}{Feng, T.}, \bibinfo{author}{Luo,
		M.}, \bibinfo{author}{Zheng, S.} \& \bibinfo{author}{Zhou, X.}
	\newblock \bibinfo{title}{Experimental demonstration of quantum finite
		automaton}.
	\newblock \emph{\bibinfo{journal}{npj Quantum Information}}
	\textbf{\bibinfo{volume}{5}}, \bibinfo{pages}{1--5} (\bibinfo{year}{2019}).
	
	\bibitem{Thesis-Feldman1998}
	\bibinfo{author}{Feldman, D.~P.}
	\newblock \emph{\bibinfo{title}{Computational mechanics of classical spin
			systems}}.
	\newblock Ph.D. thesis, \bibinfo{school}{University of California, Davis}
	(\bibinfo{year}{1998}).
	
\end{thebibliography}
\end{document}